# A standard format and a graphical user interface
# for spin system specification

A.G. Biternas[a], G.T.P. Charnock[b], Ilya Kuprov[a,*]

[a]*School of Chemistry, University of Southampton,*
*Highfield Campus, Southampton, SO17 1BJ, UK*

[b]*Department of Computer Science, University of Oxford,*
*Parks Road, Oxford, OX1 3QD, UK*

Fax:  +44 2380 594140

Email: i.kuprov@soton.ac.uk

**ABSTRACT**

We introduce a simple and general XML format for spin system description that is the result of extensive consultations within Magnetic Resonance community and unifies under one roof all major existing spin interaction specification conventions. The format is human-readable, easy to edit and easy to parse using standard XML libraries. We also describe a graphical user interface that was designed to facilitate construction and visualization of complicated spin systems. The interface is capable of generating input files for several popular spin dynamics simulation packages.

**Keywords**

## I. Introduction

The task of setting up a complicated spin system for a solid state NMR or EPR simulation is a noted test of perseverance: an aspiring theorist would find himself juggling nested time-dependent tensor rotations in half a dozen *ad hoc* conventions [1-7], struggling with Euler angle singularities [8-10] and trying to visualize interactions that occur in direct products of Lie algebras [11,12]. Function libraries [13-17], command-line [14-17] and interactive [18] *simulation* tools for spin systems are available, but convenient point-and-click *visualization and editing* tools for setting up complex calculations are in their infancy. More importantly, no standards exist (whether by ISO, IUPAC or even a consensus) on a universal spin system description format that would be applicable across all types of Magnetic Resonance spectroscopy – every major simulation package has its own spin system specification requirements. Of the existing formats, the Pople convention [19] only deals with NMR spectroscopy and the latest IUPAC recommendations only go as far as listing reasonable chemical shift and shielding tensor reporting styles [4,7]. At the time of writing, the task of setting up a complicated spin system for simulation still amounts to manual parsing of unintuitive conventions and hand-coding of the associated tensor transformations.

In this communication, we suggest a simple and general XML [20,21] format for spin system description that is the result of broad consultations within the NMR and EPR communities. The format does not attempt to introduce or change any of the current interaction specification conventions [1-4,6,7,21-26], but instead incorporates them as special cases and options into a common framework. *SpinXML* format is human-readable, extensible and easy to edit, both manually and automatically. We also describe a graphical user interface that was designed to facilitate the setting up of complicated spin systems and is capable of importing interaction data from electronic structure theory programs as well as producing input files for spin dynamics simulation packages.

## II. *SpinXML* data format

This section describes elements, types and attributes specified by the *SpinXML* schema file that is included into the Supplementary Information and available for download from the *Spinach* library website (http://spindynamics.org). An XML schema is a general description of an XML document, containing additional constraints on the structure and content of that document beyond those imposed by the syntax of XML itself [20,21]. XML has been used for a while in other areas of NMR – Agilent's VNMRJ package employs it for window layout description and an XML specification was recently proposed for phase cycles [22].

A graphical representation of the *SpinXML* schema is given in Figure 1. At the bottom of the *SpinXML* complex type (CT) hierarchy are objects intended to formalize the description of spin interaction tensors – for each interaction, amplitude and orientation information should be given. Vector and matrix complex types are not native in XML and are therefore specified explicitly as collections of double-precision real numbers. One level up, the first physically significant complex type in the *SpinXML* hierarchy is `orientation` – a property of anisotropic spin interactions that makes use of the `vector` and `matrix` CTs. Four different ways of specifying orientation are supported (Figure 1, top right corner), corresponding to the four most popular rotation conventions in Magnetic Resonance – Euler angles [23] (in degrees), angle-axis [24] (angle in degrees, unit norm vector), unit quaternion [25] and direction cosine matrix (DCM) [26]. Euler angles and quaternion specifications are simple lists of the corresponding numerical parameters, whereas DCM invokes an instance of the above mentioned `matrix` CT and angle-axis parameterization makes use of the `vector` CT for the rotation axis vector. The `SWITCH` bar that connects the four specifications indicates that only one of the four options may be invoked in each instance of the `rotation` CT. At the level of the software package making use of *SpinXML*, the parser function should be able to interpret all four rotation conventions and should be able to write at least one – from our experience working with rotation specifications in Magnetic Resonance context, we strongly recommend DCM as the default convention. *SpinXML* makes no attempt to rectify the well-documented ambiguities inherent in Euler angles [10], it only serves as a container.

At the next level in the complex type hierarchy shown in Figure 1, *SpinXML* formalizes the three general styles of spin interaction specification that are encountered in the daily practice of Magnetic Resonance spectroscopy – a scalar (isotropic interaction not requiring orientation specification), a 3×3 matrix (anisotropic interaction with orientation information already contained in the matrix) and [eigenvalue data] + [orientation data] pair. The three styles are related by a `SWITCH` bar (Figure 1 upper left corner). The scalar specification simply requires a `double`, and the matrix specification an instance of the `matrix` CT. The [eigenvalue data] + [orientation data] style includes an instance of the above mentioned `rotation` CT for the orientation information and offers the four commonly encountered ways of specifying eigenvalues: either by listing them explicitly (current IUPAC recommendation [4,7]), or by listing isotropic part, anisotropy and asymmetry [27], or isotropic part, axiality and rhombicity [2,3,28], or isotropic part, span and skew [3,29]. The mandatory attributes of the `interaction_term` CT include interaction kind (strictly from one of the following: `shielding`, `shift`, `gtensor`, `hfc`, `quadrupolar`, `exchange`, `jcoupling`, `dipolar`, `spinrotation`, `zfs`), interaction identifier (an integer), physical units and the identifier of at least one spin to which the interaction relates. The second spin (for binary interactions) and a text label are optional.

We will not discuss here the relative merits of the different styles of specifying eigenvalues – they have a long history [1-4,6,7,21-26] and a proper unification of the existing conventions is only possible in a format that includes all of them as options. This puts some strain on the software developer (a *SpinXML* parser should be able to interpret all conventions listed in Figure 1), but makes life easier for the end user. When an instance of *SpinXML* is being written rather than parsed, we would join IUPAC [4,7] in recommending the 3×3 matrix style for spin interaction tensor specification.

As a matter of practical safety, we would not recommend specifying dipolar interactions as 3×3 interaction matrices or [eigenvalue data] + [orientation data] pairs – there are quite a few papers in Magnetic Resonance literature where the listed dipole-dipole coupling constants or matrices do not correspond to a physically possible arrangement of particles in 3D space. We recommend recording inter-nuclear and inter-electron dipolar couplings by specifying particle coordinates. Electron-nuclear dipolar couplings should be supplied as anisotropic hyperfine interactions. The case of spatially proximate delocalized electrons is covered by the zero-field splitting. If dipole-dipole interactions are specified as effective spin interactions, care should be taken to ensure that the numbers provided are consistent with a physically possible set of particle coordinates.

Another problematic area is the difference between chemical shielding and chemical shift, and the associated debate [1-3] about the definition of span and skew parameters – electronic structure theory calculations report absolute nuclear shielding defined in terms of molecular energy derivatives [3], whereas experimental data is reported as fractional frequency shifts relative to a specific substance [2]. To prevent any misunderstanding SpinXML includes two types of nuclear Zeeman interaction terms: "shielding" and "shift", as well as

`reference` attribute in the `interaction_term` complex type, which is a character string that should contain the name of the reference substance. We recommend the definitions of span and skew given in the Maryland Consortium paper [1], including the subtle difference illustrated therein between the definition of tensor span for shielding and shift tensors. That having been said, although span and skew are provided as specification conventions in *SpinXML*, we would also support IUPAC [4,7] in discouraging their use – whenever possible, both chemical shift and chemical shielding should be specified using $3\times3$ interaction matrices that leave no room for ambiguity.

At the top level of the *SpinXML* format hierarchy, the `spin_system` element (Figure 1, bottom middle) contains an arbitrary number of `spin` and `interaction` elements. Each `spin` element has an integer `id`, an isotope identification string and an optional set of Cartesian coordinates. The `interaction` elements conform to the `interaction_term` complex type described in the previous paragraphs. An example of *SpinXML* specification for the spin system of $^{13}$C-labelled formaldehyde given in Figure 2 illustrates the format structure. Because of its similarity to HTML (which is actually a subset of XML), *SpinXML* syntax appears similar to a web page specification. This self-documenting property of XML [20,21] is useful because edits can be made without consulting format documentation.

Note that the isotope specification is not limited to magnetic isotopes – retaining oxygen atoms as $^{16}$O in particular is often useful in visualizations because it puts magnetic interaction schematics into a general chemical context.

## III.   Graphical user interface

A much needed stage in the spin system simulation setup process is interaction visualization. Ellipsoid plots [27,28] and spherical harmonic representations [30] of second rank tensors have been around for a while, and visualization tools dealing with subsets of spin interactions (*e.g.* Simmol [31]) are available, but a general interactive 3D GUI that would be applicable to both NMR and EPR, and be capable of exporting input files for spin dynamics simulation packages, particularly in EPR spectroscopy, has so far been missing.

Spinach GUI (designed primarily to accompany our Spinach library [17], hence the name) is an interactive 3D graphical user interface that implements all *SpinXML* features. It supports point-and-click specification of NMR and EPR spin systems, interaction tensor import from popular electronic structure theory programs (Gaussian [32], CASTEP [33], ADF [34], ORCA [35]) and export of spin system specifications into popular spin dynamics simulation packages (EasySpin [15], Spinach [17] and SIMPSON [14] at the time of writing). Import and export filters for other major programs will be added in the near future.

The main GUI window is shown in Figure 3. The atom table on the left and the interaction table on the right are self-explanatory. The central window contains a ball-and-stick representation of the molecule (bonds are drawn based on a simple distance check with a user-specified threshold) and visual representations of the following interactions:

- chemical shift and shielding tensors, using ellipsoid plots (blue by default), with ellipsoids centered at the corresponding nuclei.

- *J*-couplings, using straight lines connecting the corresponding nuclei. Interaction amplitude is mapped to the color of the line.

- hyperfine interaction tensors, using ellipsoid plots (yellow by default), with ellipsoids centered at the corresponding nuclei.

- nuclear quadrupolar interaction tensors, using ellipsoid plots (purple by default), with ellipsoids centered at the corresponding nuclei.

Ellipsoid plots are generated in the standard way [25,31]: a unit sphere is scaled by the moduli of the eigenvalues in the three primary directions, rotated into the principal axis frame of the tensor and translated to the point of the corresponding nucleus. Blue axes are drawn inside for positive eigenvalues and red axes for negative eigenvalues.

Dipolar interaction tensors are not visualized – inter-nuclear dipolar coupling is visually apparent from the distances and electron-nuclear dipolar coupling is contained in the hyperfine interaction. In systems with multiple electrons, the inter-electron dipolar coupling is either contained in the distances (in the individual electron spin representation) or in the zero-field splitting tensor (in the total electron spin representation).

It is often the case in Magnetic Resonance simulations that electrons do not have specific Cartesian coordinates, being instead delocalized over the nuclear ensemble and manifesting themselves through hyperfine interactions. For this reason electrons are drawn separately in the lower part of the central area in. Electron interaction ellipsoids rotate synchronously with the rest of the molecule, but the electrons themselves (visualized as translucent blobs) do not move around the visualization window. Zero-field splitting tensors and *g*-tensors are visualized as ellipsoids centered on their corresponding electrons and inter-electron exchange couplings are shown as coils with the interaction amplitude mapped to the color.

A summary of the visualization methods is given in Table 1. *Visualization* tab in the upper part of the main window controls the appearance and scaling of the ellipsoids as well as magnitude-color maps in the 3D view using logarithmic sliders. Visualization of individual interactions may be switched on and off using the tick boxes. *NMR* and *EPR* buttons switch the 3D view to the visualization of the corresponding interactions – shielding, shift, *J*-coupling, quadrupolar coupling for the NMR mode; *g*-tensor, hyperfine coupling, exchange coupling, zero-field splitting for the EPR mode.

The primary format for spin system data storage and retrieval is *SpinXML*, but the GUI can also import Gaussian 03/09 logs (*.log, *.out), Cartesian XYZ files (*.xyz, coordinates only, isotopes are guessed) and both versions of CASTEP MagRes files (*.magres). When multiple instances of the relevant tables are present in the file (*e.g.* multiple coordinate sections in geometry optimizations), the last section is read. For Gaussian 03/09 calculations, the detailed printing option is required in the route section of the input file.

Electronic structure theory calculations often produce large quantities of small interactions (*e.g. J*-couplings between remote spins) that are inconsequential in practical simulations. At the point of data import the GUI offers an option to ignore interactions with total magnitude (defined as the Frobenius norm of the corresponding tensor) below the user-specified value.

The 3D view is rendered using the OpenGL library [36]. Real-time rotations are implemented using the ARCBALL scheme [37] that assumes the mouse to be moving on the surface of a ball centered on the model. Dragging the pointer forms an arc that the system is rotated along. When the pointer is dragged outside the ball (*e.g.* at the edge of the 3D view panel), the model is rotated only around the axis perpendicular to the screen. The 3D view is cross-referenced with both tables – when an atom is selected in the 3D view, its coordinate line in the atom table is highlighted in blue and its associated spin interactions in the interaction table are highlighted in yellow.

The *Interactions* table on the right side of the main window provides a list of all spin interaction present in the system, except for the dipole-dipole couplings that are controlled *via* Cartesian coordinates in the left hand side table. For each interaction, a unique numerical ID, a user-specified label, the IDs of the participating spins and the type of the interaction may be edited directly in the table. Eigenvalues and orientation may be edited by pressing "Edit" in the table and making changes in the *Magnitude & Orientation* dialogue window shown in Figure 4. The GUI offers five ways to edit an interaction. The user can change the interaction matrix, eigenvalues, spherical tensor coefficients, Euler angles, or angle-axis rotation angle. If any of those are changed, the content of the entire window is recomputed to reflect the changes and the 3D view is updated accordingly. In the cases where manual edits have the potential to violate a convention (*e.g.* break the norm of a directional cosine matrix or a quaternion), direct edits are disabled and the corresponding fields are greyed out, they are only updated in response to convention-preserving edits. The flowchart of rotational convention updates is given in Figure 5.

The interface to spin dynamics simulation packages follows the same design philosophy as the very successful Gaussian/GaussView [32] pair of programs in electronic structure theory. An example of the export dialogue window is shown in Figure 5. The GUI currently generates ASCII text files containing spin system description inputs for EasySpin [15], Spinach [17] and SIMPSON [14] packages with support for other major simulation programs currently in the works. Only the spin system description part is generated: *spinsys* section of SIMPSON input and the corresponding Matlab code for Spinach and EasySpin packages – experiment description parts should be appended to the resulting text file by the user.

## III.    Conclusions and outlook

SpinXML unifies under one roof all major spin interaction specification conventions and provides a way of tying them up into a multi-spin system description format that is applicable to both NMR and EPR spectroscopy. The format is based on the industry standard XML markup language and benefits from the existence of standard validation, generation and parsing tools in all major programming languages. It is our hope that it would facilitate the storage and exchange of spin system data, particularly with the recently created protein-scale simulation tools [17]. The associated graphical user interface provides a user-friendly way of setting up complicated spin systems as well as a convenient way of importing magnetic interaction data from electronic structure theory packages.

## Acknowledgements

We are grateful to Marina Carravetta, Jean-Nicolas Dumez, Robin Harris, Paul Hodgkinson, Peter Hore, Edmund Howard, Malcolm Levitt, Ivan Maximov, Niels Christian Nielsen, Konstantin Pervushin, Giuseppe Pileio, Vadim Slynko, Zdenek Tosner, and Thomas Vosegaard for useful feedback during SpinXML and GUI development. This project is supported by EPSRC (EP/F065205/1, EP/H003789/1).

**Figure captions**

**Figure 1.** A visual representation of the *SpinXML* format schema. The four fundamental complex types, described in detail in the main text, are `vector`, `matrix`, `rotation` and `interaction_term`. The `spin_system` XML element contains an arbitrary number of `spin` sub-elements and an arbitrary number of `interaction` sub-elements, each of which must conform to the `interaction_term` complex type. Dashed lines indicate optional attributes. An example of the XML specification conforming to this schema is given in Figure 2. The schema file is included in the Supplementary Information.

**Figure 2.** An example of a SpinXML file for the spin system of $^{13}$C-labelled formaldehyde. Atomic coordinates, chemical shielding tensors and *J*-couplings were imported from a GIAO DFT M06/cc-pVTZ calculation in Gaussian09 [32].

**Figure 3.** A screenshot of the main window of the graphical user interface. **(1)** Nuclei and unpaired electrons list. Cartesian coordinates are displayed for every nucleus and unpaired electron, and isotope mass numbers for each nucleus. **(2)** Interaction visualization control panel, containing logarithmic (shielding, hyperfine, quadrupolar, g-tensor, zero-field splitting) and linear (exchange, J-couplings) scaling sliders for visual representations of the interactions. **(3)** Electrons area, showing *g*-tensors (ellipsoids), zero-field splitting tensors (ellipsoids) and exchange couplings (spirals). **(4)** Nuclei area, showing shielding tensors (ellipsoids), hyperfine tensors (ellipsoids), *J*-couplings (lines, drawn if interaction exceeds a set threshold) and quadrupolar tensors (ellipsoids). **(5)** Interaction list, containing basic interaction information and editing buttons. **(6)** Selected nucleus, showing its identification number. The corresponding table rows are highlighted in pink on the left hand side (coordinates and isotope information) and yellow on the right hand side (interactions).

**Figure 4.** Interaction editing dialogue window, accessed by clicking "Edit" in the right hand side table of the main window. **(1)** Interaction tensor matrix (type and units are indicated above). **(2)** Eigenvalues of the interaction matrix. **(3)** Eigenvectors of the interaction matrix. (**4**) Euler angles connecting the principal axis frame of the interaction tensor to the laboratory frame of reference. (**5**) Interaction axiality, rhombicity, span and skew. (**6**) Irreducible spherical components of the interaction tensor. (**7**) Angle-axis rotation parameters connecting the principal axis frame of the interaction tensor to the laboratory frame of reference. (**8**) Quaternion rotation parameters connecting the principal axis frame of the interaction tensor to the laboratory frame of reference. (**9**) Wigner rotation matrix connecting the principal axis frame of the interaction tensor to the laboratory frame of reference.

**Figure 5.** A map of amplitude and orientation information conversion within the interaction editing dialogue window. Double arrows indicate two-way conversion (when one side is edited, the other side is updated). Single arrows indicate one-way conversion (the receiving side is updated, but is not itself editable). Green boxes indicate representations that typically arrive from electronic structure theory programs, green and blue boxes indicate representations that can be imported from *SpinXML*, grey boxes indicate significant but uncommon representations that are calculated and reported within the GUI for diagnostic purposes.

**Figure 6.** An example of a simulation package export dialogue window producing an ASCII text input file for a third-party spin dynamics simulation package. EasySpin [15] export window is shown; the GUI also supports Spinach [17] and SIMPSON export [14]. (**1**) Route specification area, switching the simulation between liquid state, slow motion and solid state in the case of EasySpin. (**2**) Case-specific and program-specific parameter specification area. (**3**) Text output preview window.

**Table 1.** Interaction visualization methods and default units.

| Interaction | Visualization method | Default units in GUI |
|---|---|---|
| hyperfine interaction | ellipsoid on the nucleus | Gauss |
| chemical shielding and shift | ellipsoid on the nucleus | ppm |
| quadrupolar interaction | ellipsoid on the nucleus | MHz |
| *J*-coupling | coloured line between nuclei | Hz |
| dipolar coupling (=distances) | particle positions | Ångstrom |
| *g*-tensor | ellipsoid on the electron | dimensionless |
| zero-field splitting | ellipsoid on the electron | MHz |
| exchange interaction | coiled line between electrons | MHz |

**Figure 1**

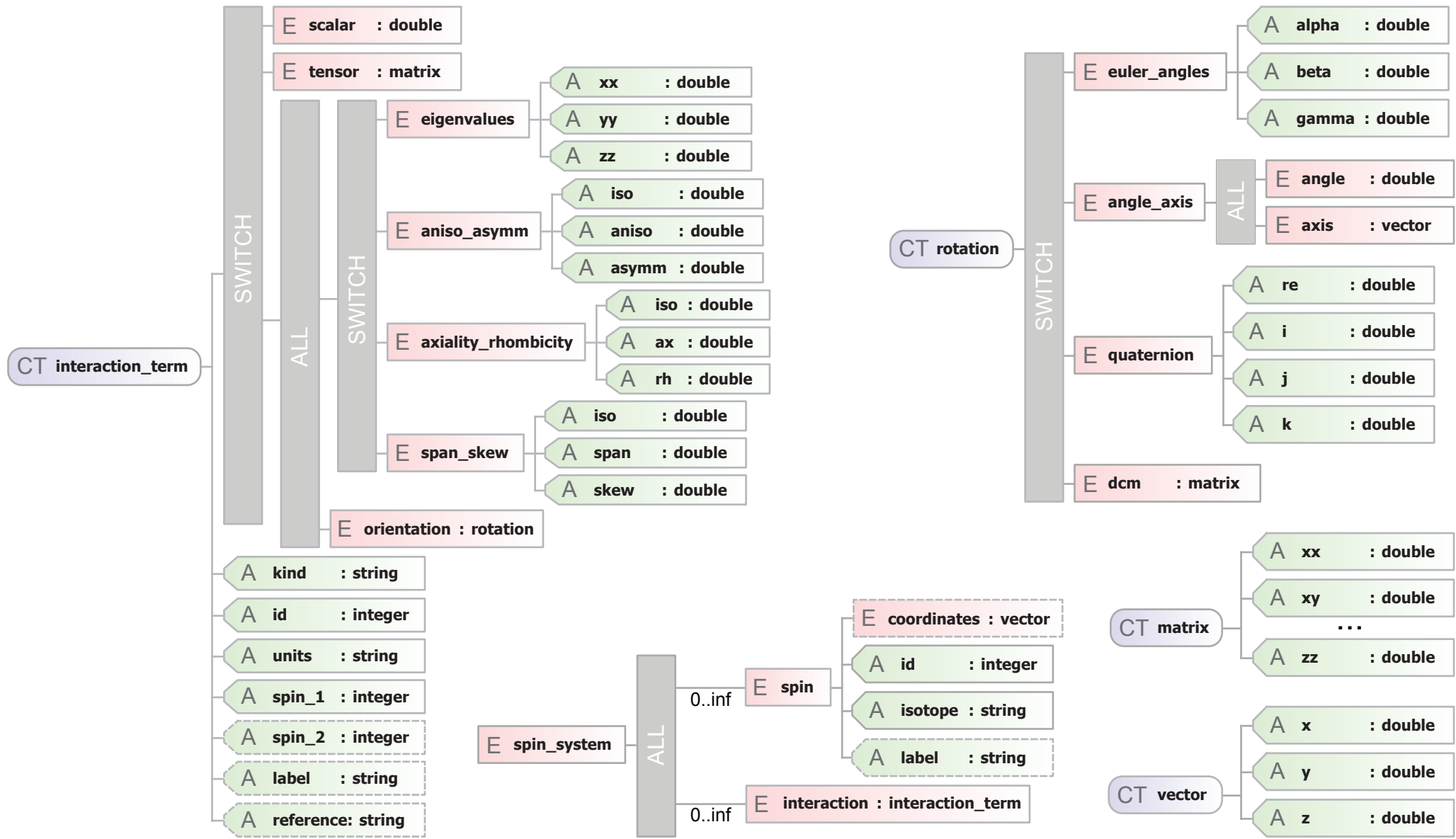

**Figure 2**

```xml
<spin_system>

  <spin number="1" isotope="1H" label="Proton A" >
    <coordinates x="0.937"  y="0.000" z="0.000" />
  </spin>

  <spin number="2" isotope="1H" label="Proton B" >
    <coordinates x="-0.937" y="0.000" z="0.000" />
  </spin>

  <spin number="3" isotope="13C" label="Carbon" >
    <coordinates x="0.000" y="-0.526" z="0.000" />
  </spin>

  <spin number="4" isotope="16O" label="Oxygen" >
    <coordinates x="0.000" y="0.673" z="0.000" />
  </spin>

  <interaction kind="shielding" units="ppm" spin_1="1" reference="absolute">
    <eigenvalues xx="20.2" yy="21.8" zz="22.2" />
    <rotation>
      <euler_angles alpha="230.4" beta="0.0" gamma="0.0" />
    </rotation>
  </interaction>

  <interaction kind="shielding" units="ppm" spin_1="2" reference="absolute">
    <tensor xx="21.16" xy="-0.76" xz="0.00"
            yx="-0.76" yy="20.87" yz="0.00"
            zx="0.00"  zy="0.00"  zz="22.18" />
  </interaction>

  <interaction kind="shielding" units="ppm" spin_1="3" reference="absolute">
    <span_skew iso="-25.31" span="214.70" skew="0.135" />
    <rotation>
      <euler_angles alpha="180" beta="0.0" gamma="0.0" />
    </rotation>
  </interaction>

  <interaction kind=jcoupling" units="Hz" spin_1="1" spin_2="2" >
    <scalar>29.13</scalar>
  </interaction>

  <interaction kind=jcoupling" units="Hz" spin_1="1" spin_2="3" >
    <scalar>256.9</scalar>
  </interaction>

  <interaction kind=jcoupling" units="Hz" spin_1="2" spin_2="3" >
    <scalar>256.9</scalar>
  </interaction>

</spin_system>
```

**Figure 3**

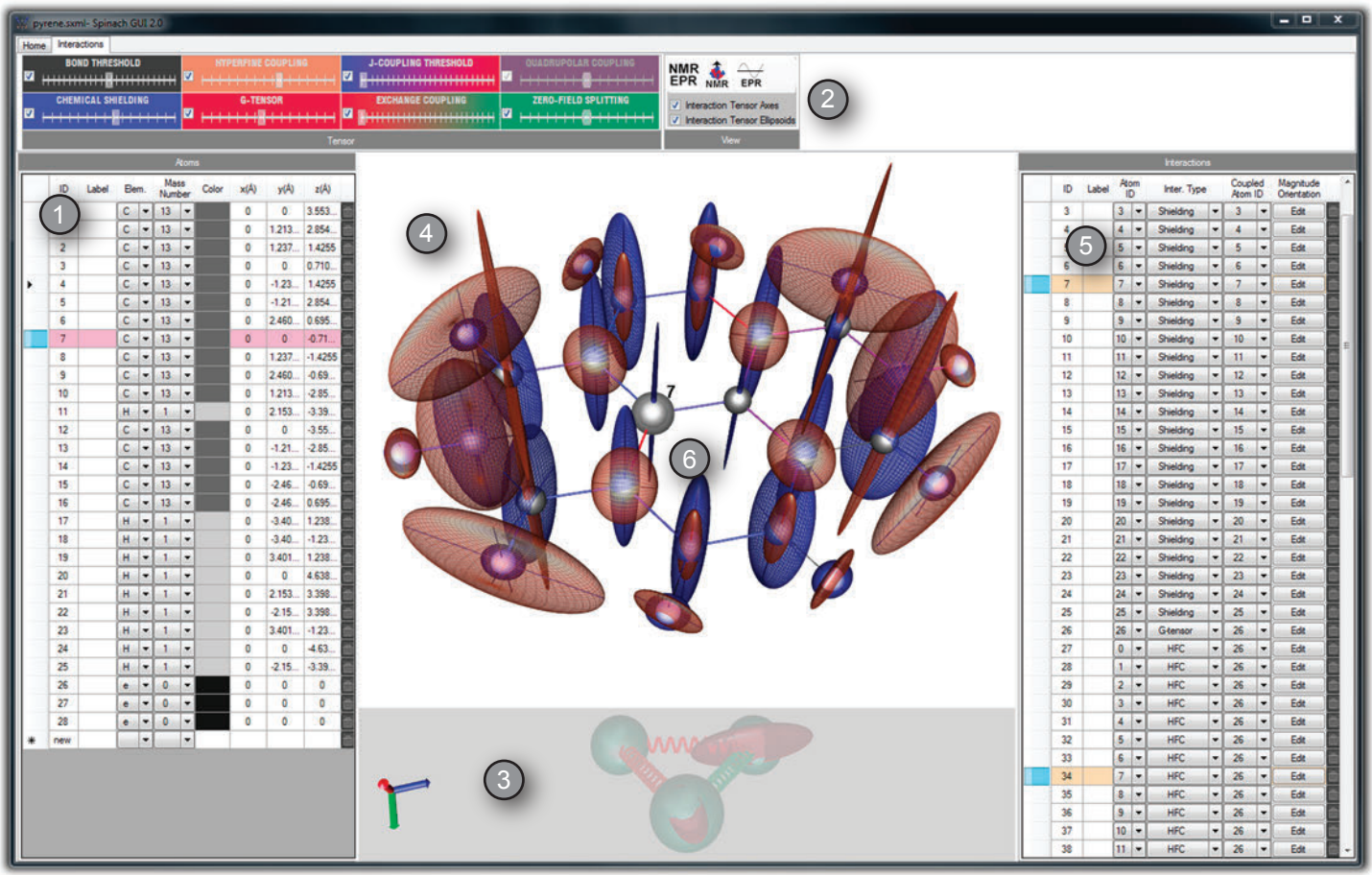

**Figure 4**

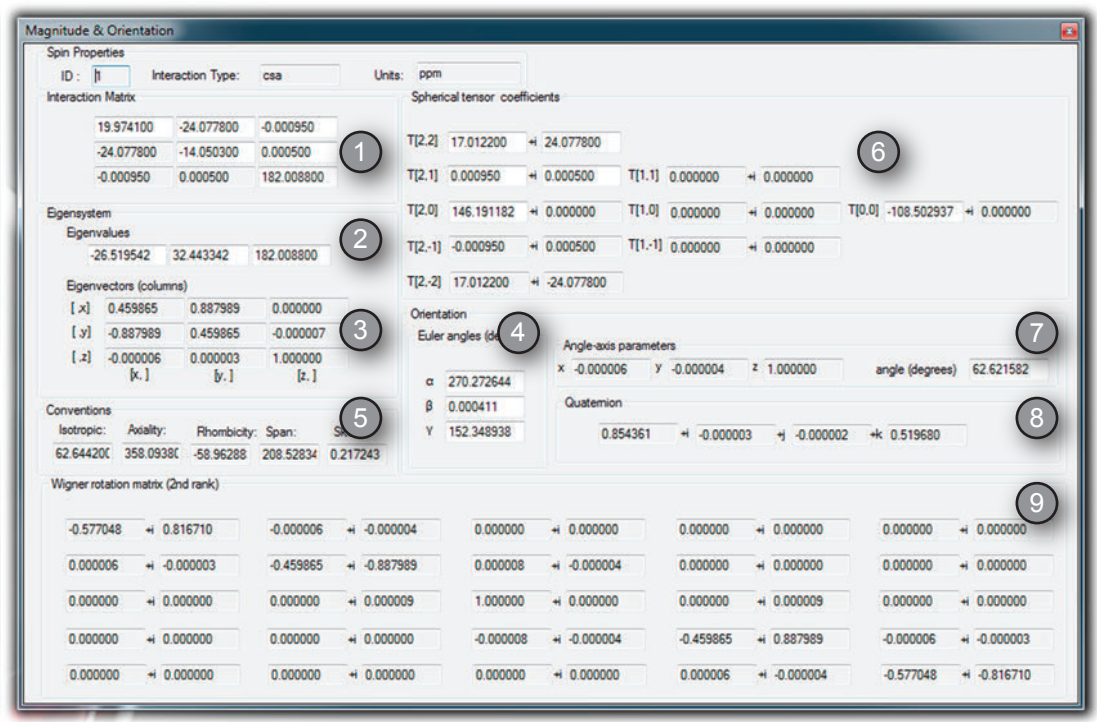

**Figure 5**

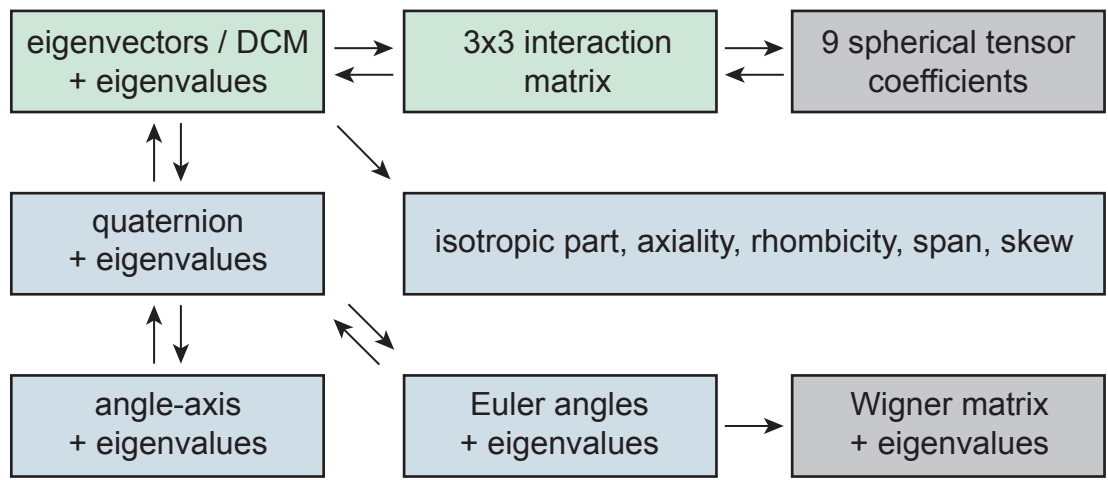

**Figure 6**